\documentclass[english,twocolumn]{revtex4-2}
\usepackage[T1]{fontenc}
\usepackage[latin9]{inputenc}
\usepackage{geometry}
\usepackage{verbatim}
\usepackage{mathrsfs}
\usepackage{graphicx}
\usepackage{esint}
\usepackage{babel}
\begin{document}
\title{On the dynamics of convolutional recurrent neural networks near their
critical point}
\author{Aditi Chandra$^{12}$ and Marcelo O. Magnasco$^{1}$}
\affiliation{$^{1}$Rockefeller University, New York. }
\affiliation{$^{2}$Physics Department, University of Oxford, UK}
\begin{abstract}
We examine the dynamical properties of a single-layer convolutional
recurrent network with a smooth sigmoidal activation function, for
small values of the inputs and when the convolution kernel is unitary,
so all eigenvalues lie exactly at the unit circle. Such networks have
a variety of hallmark properties: the outputs depend on the inputs
via compressive nonlinearities such as cubic roots, and both the \emph{timescales
of relaxation} and the \emph{length-scales of signal propagation}
depend sensitively on the inputs as power laws, both diverging as
the input $\to0$. The basic dynamical mechanism is that inputs to
the network generate ongoing activity, which in turn controls how
additional inputs or signals propagate spatially or attenuate in time.
We present analytical solutions for the steady states when the network
is forced with a single oscillation and when a background value creates
a steady state of ``ongoing activity'', and derive the relationships
shaping the value of the temporal decay and spatial propagation length
as a function of this background value. 
\end{abstract}
\maketitle

\section{Background}

Recurrent neural networks (RNNs) are famously difficult \citep{vanish1,rnn1,rnn2}.
While they are Turing universal \citep{turing0,turing2,turing3},
their full potency has been hard to harness due to their complexity.
RNNs are difficult to analyze, or predict more than a few steps into
the future, and they are difficult to train due to the vanishing gradient
problem \citep{vanish1,rnn1} and related issues. So at present we
\textbf{neither} have a clear programming paradigm to \textbf{design}
them, nor a generally effective method to \textbf{train} them. 

Meanwhile neurobiology confronts us with networks that are almost
always recurrent, often heavily so. %
{} The need to understand recurrent biological networks has spawned
considerable efforts to advance their study. Here we undertake an
in-depth description of a very small and simple subset of all RNNs,
in the hope it may serve as a beachhead into more general cases. 

A concept often discussed in computational neuroscience is \emph{criticality}.
One category is \emph{dynamical criticality}, where the real parts
of stability parameters approach zero. Two early models are the line
attractors \citep{seung}, where one direction acquires a zero eigenvalue,
and the Hopf bifurcation scenario for hair-cell dynamics \citep{hairbundle,essential,tuningcurve},
where the eigenvalues are purely imaginary. Since then, there's been
continued interest in this scenario, both theoretically \citep{antihebbian,waveatten,waveatt2,chaos}
as well as experimentally \citep{ecog1,ecog2,ecog3,ecog4}, and it's
been conjectured to be involved in dynamical reconfiguration and functional
flexibility \citep{currentopinion,entropy}. Furthermore, it has been
argued that traveling waves play an essential role in the theory of
recurrent neural networks. both in terms of maintaining the ``working
memory'' of the RNN \citep{waves1} as well as transferring information
between working variables \citep{waves2}. Traveling waves are sustained
by dynamically critical eigenmodes. 

While many models of neuronal function are written as continuous-time
differential equations, artificial neural networks are usually expressed
by discrete-time recurrences. Moving from continuous- to discrete-time
requires an exponentiation operation: purely imaginary (critical)
eigenvalues are mapped by the exponential to lay on the unit circle
\citep{strogatz}, and connectivities are mapped to orthogonal or
unitary matrices; see \citep{chaos,arxiv} for an explicit derivation
applicable to our case. There has been a lot of recent interest from
the ML community in recurrent networks with either orthogonal or unitary
evolutions \citep{uni1,uni2,uni3,uni4}; as maintaining orthogonality
is algorithmically difficult (a.k.a. the parametrization of the Stieffel
manifold), even initializing the weights to be orthogonal or unitary
helps with subsequent evolution of the training \citep{init1,init2,init3}.
Here we will explore a model for which establishing and maintaining
unitarity is algorithmically fast \citep{arxiv}. 

\section{The Model}

\begin{figure}[tb]

\includegraphics[width=8.4cm]{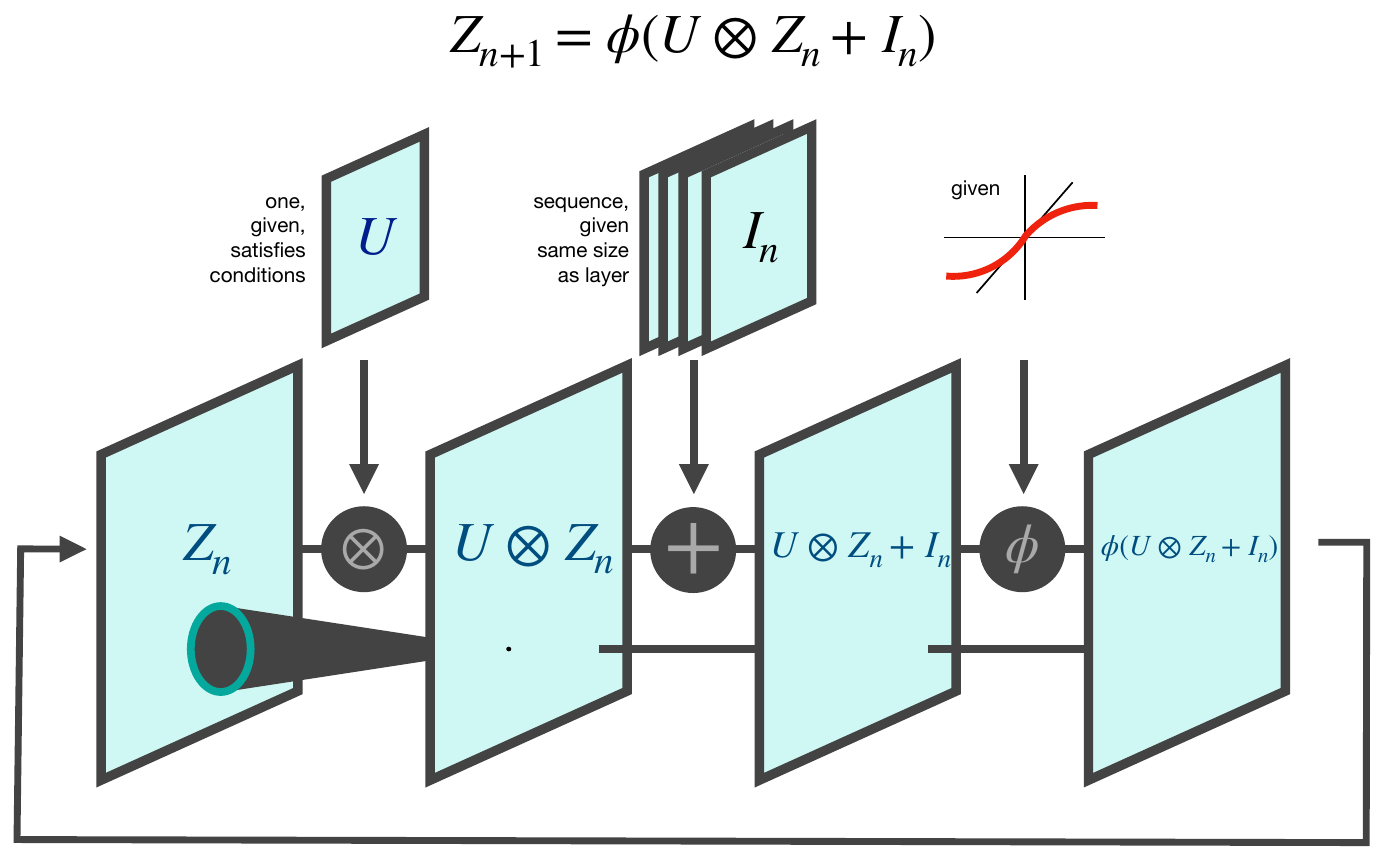}\caption{Schematic of the map Eq \ref{eq:model}. For any given element in
the layer, all the inputs are added together; the inputs from other
elements in the layer is given by the convolution, and the external
input is added element-wise. Finally the activation function is applied,
and the result becomes the values of the layer for the next time-step.
\label{fig:Schematic}}

\end{figure}

A complex-valued layer $Z$ whose structure supports convolutions
(e.g.square lattices in $D$ dimensions with periodic boundary conditions)
evolves through the discrete-time recurrent dynamics

\begin{equation}
Z_{n+1}=\phi\left(U\otimes Z_{n}+I_{n}\right)\label{eq:model}
\end{equation}
where $Z_{n}$ denotes the values of the layer at time $n$, $U$
is a convolution kernel, $\otimes$ is the convolution operation native
to $Z$, and the activation function $\phi$ is an \emph{element-wise
}function given by 
\begin{equation}
\phi(z)=\frac{z}{\sqrt{1+|z|^{2}}}\label{eq:phi}
\end{equation}
We will want a convolution kernel $U$ whose action, when viewed as
a linear operator, is \emph{unitary}. A computationally fast way to
generate unitary convolutions \citep{arxiv} is to use the \emph{convolutional
exponential }$e_{\otimes}^{A}$ applied to a anti-Hermitian kernel
$A$; this can be explicitly computed in Fourier space as
\begin{equation}
U=e_{\otimes}^{A}\equiv\mathscr{F}^{-1}\left[\exp\left(\mathscr{F}\left[A\right]\right)\right]\quad{\rm with}\quad A^{\dagger}=-A\label{eq:convexpo}
\end{equation}
with the exponentiation in the rhs taken element-wise. The anti-Hermitian
kernel $A$ reverses sign upon flipping centrally and complex-conjugating.
See Figure \ref{fig:Schematic}The convolutional exponential is a
fast $N\log N$ operation. All eigenvalues of a unitary operator have
absolute value equal to 1 and therefore the operator does not expand
or contract any direction; in dynamical systems parlance, it satisfies
Liouville's theorem and preserves phase space volume, with expansions
or contractions only coming locally from the slope of $\phi$. In
fact with these choices of activation and kernel, Eq. \ref{eq:model}
is time-reversible. 

In the theory of ML, equation \ref{eq:model} is a convolutional recurrent
neural net, where the emphasis is on training the weights $U$ (or
better stated, $A$ if we want to keep the system unitary) while in
the theory of dynamical systems it is a \emph{coupled map lattice
}\citep{cml1,cml2}, where the emphasis is on complex asymptotic behavior
upon use of chaotic maps $\phi$. 

\subsection{Notation}

In the rest of this paper we shall (ab)use the following notation
introduced in \citep{arxiv}, which is regrettably necessary as it
is very easy to get confused. A one megapixel image, despite being
arranged in 2D, is \textbf{not actually a matrix} because we do not
rely on multiplying two images together using matrix multiplication
-- using the dot products of the first one's rows with the second
one's columns. It is a 2D \emph{array}, but \emph{not} an actual \emph{matrix}
embodying a linear algebra operator over a vector space. For our analysis
it is a \emph{vector} with 1 million elements, an element of the vector
space of all possible images. Similarly, a 2D convolution kernel looks
like a matrix but it actually isn't. It is an element of the vector
space of kernels, and possesses a central element, which we will call
$(0,0)$, which is the kernel element that gets multiplied by \emph{this
}pixel when convolving. We'll use Python notation in referring to
elements to the left and below this center as having negative indices.
OTOH the operation of convolving with a kernel $K$, denoted abstractly
as $K\otimes$, is indeed a proper operator in the sense of linear
algebra, and hence it is (abstractly) a matrix if we were to express
it in any given basis; it is a matrix of 1 million rows times 1 million
columns with a characteristically repetitive structure and lots of
empty space, so we never actually write it out that way because it
is wasteful, but that is what it is. 

\begin{figure}
\includegraphics[width=8.6cm]{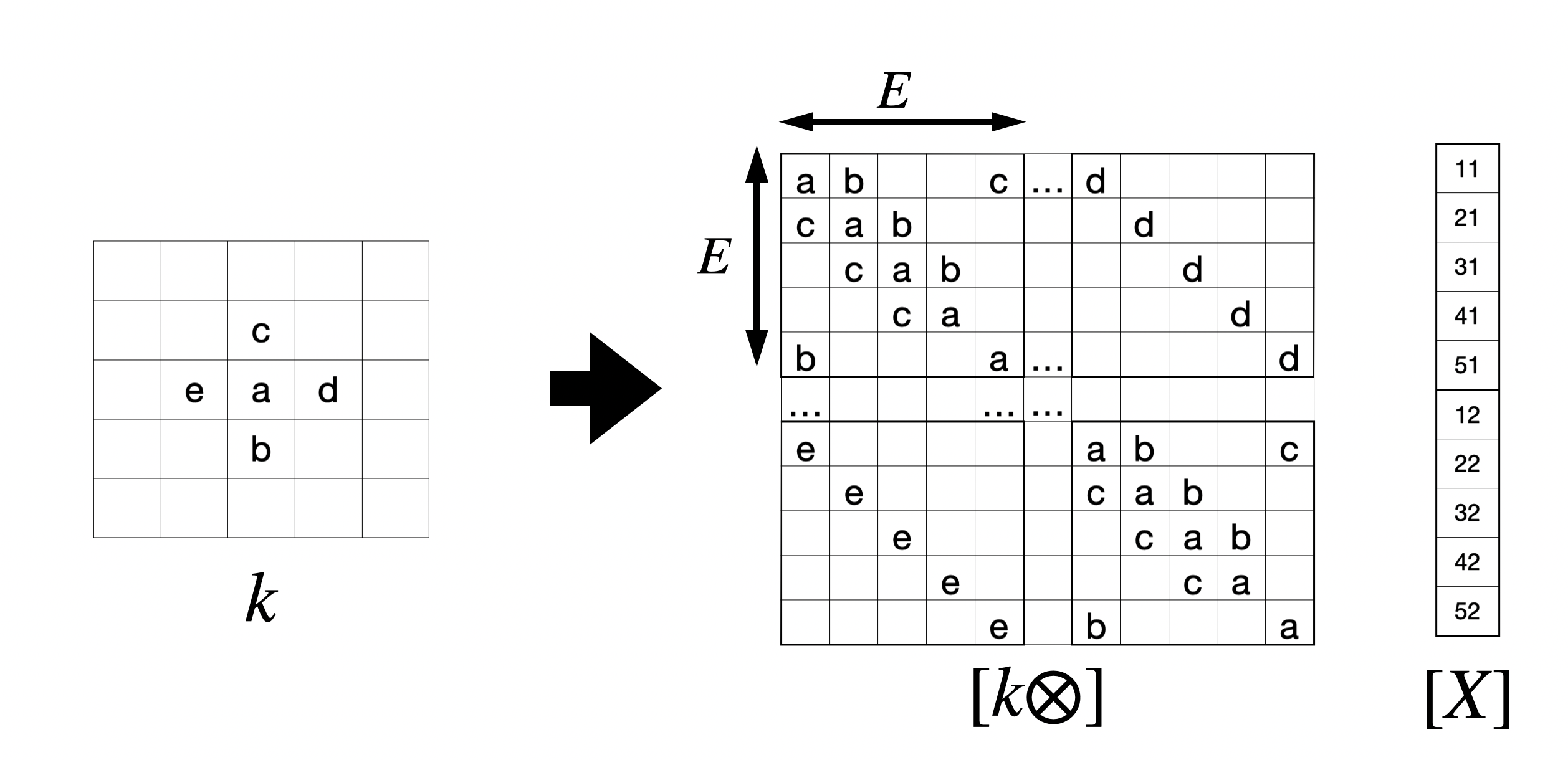}\caption{Choose as a basis the \textquotedblleft reshape\textquotedblright{}
operation from an $E\times E$ square to a vector of length $E^{2}$
$(0\le i<E,0\le j<E)\ \to\ k\equiv i+Ej\ <E^{2}$ and a convolution
which for every point in the lattice adds up the first neighbors with
coefficients $y_{ij}=ax_{ij}+bx_{i+1,j}+cx_{i-1,j}+dx_{ij+1}+ex_{ij-1}\qquad\forall ij$.
This operation maps the kernel onto a sparse array with lots of repetitive
diagonal structures and off-diagonal stuff for boundary conditions.
Any matrix property invariant under change of basis is a property
of the kernel too. For example, transposition -> central symmetry.
\label{fig:reshape}}
\end{figure}

We shall denote by $Z$ and $U$ the layer and the kernel in their
``natural'' topology, e.g. as 1D or 2D arrays. However, it will
be useful throughout to choose a basis for the vector space of all
$Z$s, and we shall denote by $\left[Z\right]_{B}$ the representation
of $Z$ in the basis $B$; this representation is always a vector
regardless of the dimensionality of $Z$. The dimension of this representation
is the total number of elements in the layer: the vector representing
a 1D layer of size $N$ has $N$ elements and looks the same as the
original, and a 2D layer of width $W$ and height $H$ has $W\times H$
elements. There is one canonical basis for $Z$ in 1D: the layer itself
laid out as a vector. In 2D, there are two canonical bases: the ``Fortran''
or ``column-major'' basis, which takes each column of $Z$ as a
vector and concatenates them sequentially, or the ``C'' or ``row-major''
basis, where it is now the rows who are concatenated one after the
other. We declare ourselves agnostic on this quasi-religious issue.
The other basis that we shall need to use is the Fourier basis, which
we shall call $\mathscr{F}$, and which will denote Fourier transformation
in the dimension appropriate to the layer and in the same major order
that is used for the layer. 

The operation of convolving an object with some kernel $K$, which
we could call $K\otimes$ (``$K$ convolved with'') is a linear
operator; $\left[K\otimes\right]_{B}$ is the representation of this
linear operator on the basis $B$, and it is a \emph{square matrix}
whose sides are equal to the total number of elements of $Z$. See
Figure \ref{fig:reshape}. In the 1D case, it is an $N\times N$ matrix,
in the 2D case a $WD\times WD$ matrix. Using this notation, 
\[
\left[K\otimes Z\right]=\left[K\otimes\right]\times\left[Z\right]
\]
where $\times$ is matrix multiplication: $\left[K\otimes Z\right]$
is a vector obtained by matrix multiplication of the matrix $\left[K\otimes\right]$
times the vector $[Z]$. It is computationally inefficient to do it
this way, but this what the convolution is. Any matrix property of
$\left[K\otimes\right]$ which is invariant under change of basis,
e.g. its eigenvalues, can therefore be properly attributed to $K$
itself, and so henceforth we will refer to ``the eigenvalues of $K$''
meaning ${\rm eig}\left(\left[K\otimes\right]\right)$; under no circumstance
should it be interpreted as the eigenvalues of the array where $K$
is stored. In addition, the operation of flipping $K$ ``antipodally''
through its center (element ``(0,0)'', the element that multiplies
the same pixel we are assigning to in the convolution), call it $K^{F}$
corresponds, in any dimension, to transposing $\left[K\otimes\right]$,
thus allowing us to use the words ``symmetric'' or ``Hermitian''
as applied to $K$: it is not a transpose of the array storing $K$
but rather a central flip swapping each element with its antipode.
Thus $\left[K^{F}\otimes\right]=[K\otimes]^{\top}$. 

Since matrix multiplication of the representation is the representation
of the application of the convolution it follows that
\[
[(K\otimes K)\otimes]=[K\otimes]\times[K\otimes]
\]
and so on for all powers, from where, by summing order by order the
Taylor series of the matrix exponential, 

\[
e^{[K\otimes]}=I+[K\otimes]+\frac{[K\otimes]^{2}}{2!}+\frac{[K\otimes]^{3}}{3!}+\cdots
\]
we define $e_{\otimes}^{K}$, the \emph{convolutional exponential}
of $K$ as the kernel whose representation equals the matrix exponential
of the representation of $K\otimes$:

\[
[e_{\otimes}^{K}]\equiv e^{[K\otimes]}
\]
from where 

\[
e_{\otimes}^{K}\equiv1+K+\frac{K\otimes K}{2!}+\frac{K\otimes K\otimes K}{3!}+\cdots
\]
Recurrent use of the Fourier convolution theorem, $K\otimes K=\mathscr{F}^{-1}[\mathscr{F}[K]^{2}]$,
gives us $K\otimes K\otimes K=\mathscr{F}^{-1}[\mathscr{F}[K]^{3}]$
and so on and so forth, and summing all orders gives Eq. \ref{eq:convexpo}. 

Therefore, in order to generate a unitary kernel, we take an arbitrary
\emph{real} kernel $K$ and 
\begin{equation}
K\ \to\ \ \ A=\frac{K-K^{F}}{2}+i\frac{K+K^{F}}{2}\ \ \ \to\ U=e_{\otimes}^{A}\label{eq:k_to_a_to_u}
\end{equation}
with the first step being fully invertible. 

\section{Basic analysis}

A number of basic considerations were elaborated in \citep{arxiv}
which we will not repeat here; we add here what we need for this Paper. 

\subsection{Choice of activation function $\phi$}

This model is based on a discrete-time dynamics; and correspondingly
we have chosen the activation function $\phi$ to correspond to the
discrete-time dynamics of a marginally-stable fixed point. To derive
$\phi$, consider the equation $\dot{z}=-|z|^{2}z$. Being rotationally
invariant we can assume $z$ to be real and positive. Being 1D it
admits a closed-form solution in quadratures:
\[
\frac{dz}{-z^{3}}=dt\quad\to\quad\int_{z(0)}^{z(t)}\frac{dz}{-z^{3}}=\int_{0}^{t}dt\to
\]
\[
\to\quad\left(z(t)^{-2}-z(0)^{-2}\right)=3t
\]
from where we obtain the solution and cast it as a flow $\phi_{t}$:
\[
z(t)=\frac{z(0)}{\sqrt{1+3tz(0)^{2}}}\quad\to\quad\phi_{t}(z)\doteq\frac{z}{\sqrt{1+3t\left|z\right|^{2}}}
\]
Being a flow, i.e. the outcome of integrating an ODE, it comes equipped
with a \emph{group structure}: evolving to $t_{1}+t_{2}$ is the same
as evolving to $t_{1}$ and then evolving $t_{2}$ more: 
\begin{equation}
\phi_{t_{1}}\circ\phi_{t_{2}}=\phi_{t_{1}+t_{2}}\label{eq:groupstructure}
\end{equation}
and is invertible: 
\[
\phi_{t}^{-1}=\phi_{-t}=\frac{z}{\sqrt{1-3t\left|z\right|^{2}}}
\]
Finally, $\phi$ and rotations commute: 
\begin{equation}
\phi(e^{i\theta}z)=e^{i\theta}\phi(z)\label{eq:rotinv}
\end{equation}
 so $\phi$ gives us the exact analytical solution to the (unforced)
Hopf normal form.

In the remainder of this paper we choose $t=1/3$ and we drop the
time index. 

\subsection{Compression and singularity}

\begin{figure}
\includegraphics[height=4cm]{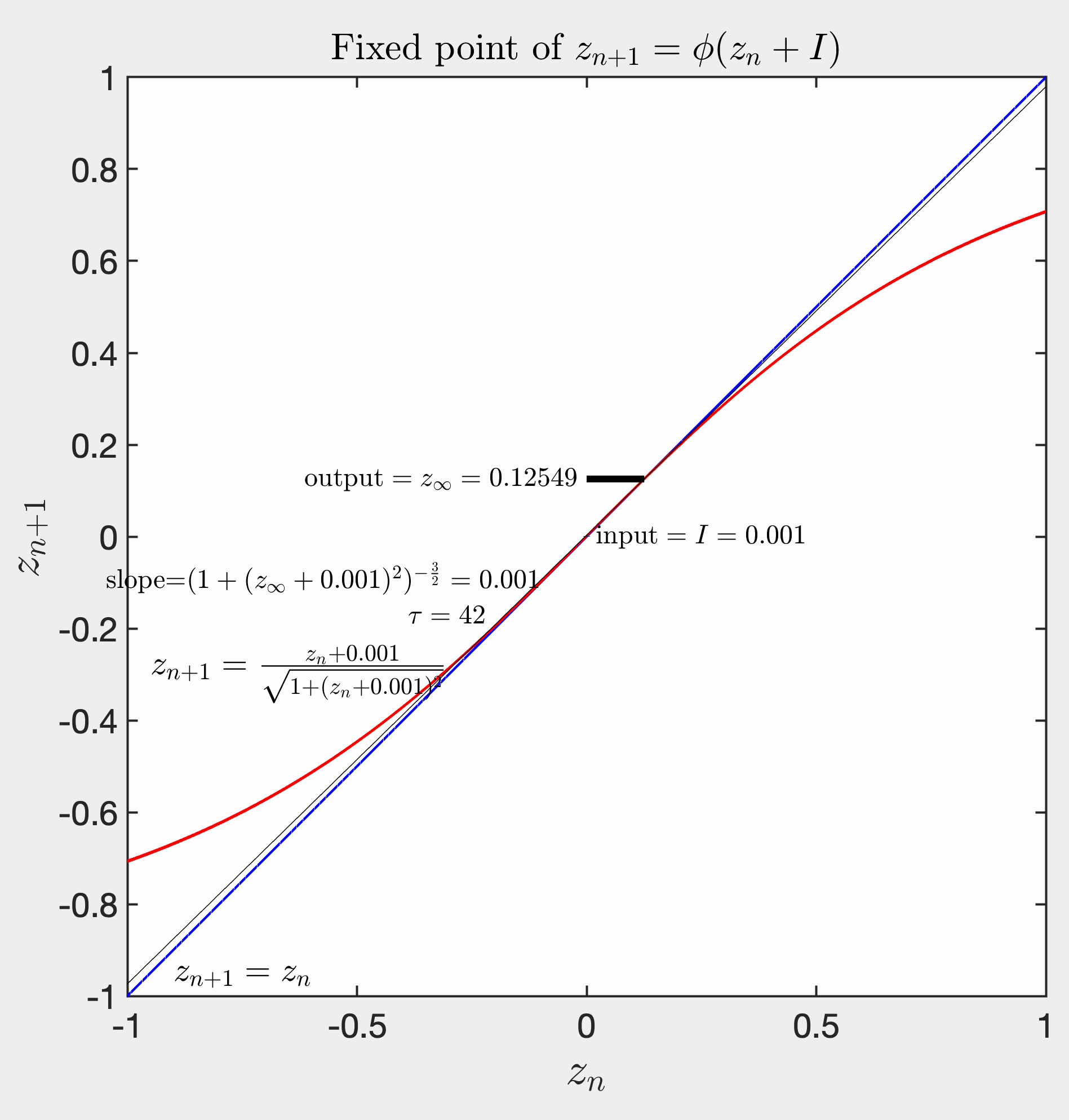}\includegraphics[height=4cm]{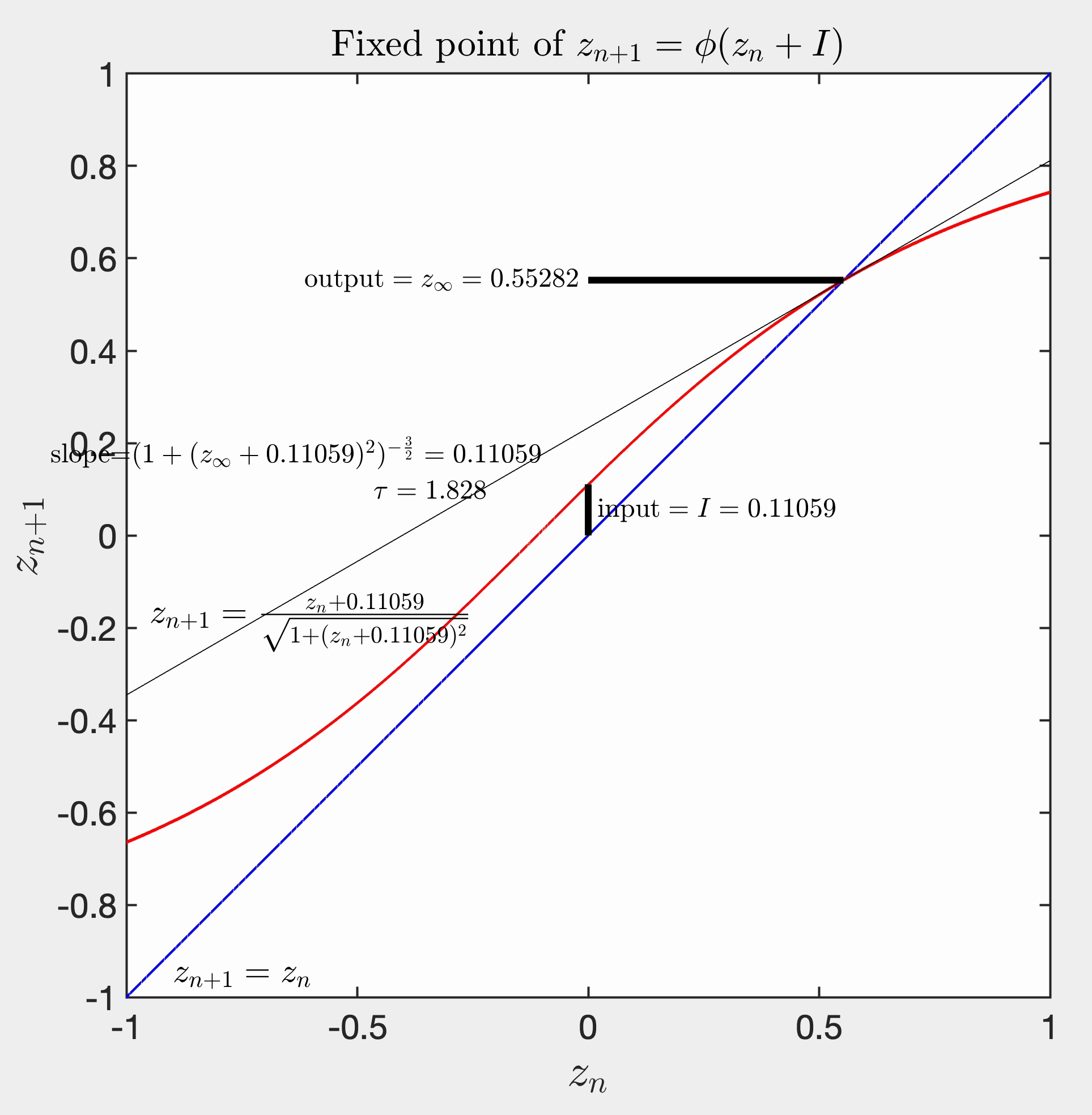}\caption{Fixed point and slope at the fixed point. In these diagrams, the fixed
point lies at the intersection of the curve with the diagonal; as
the input displaces the curve upwards, the fixed point moves to the
right. Left, with a very small input (0.001) representing a nearly
invisible displacement of the curve with respect to the diagonal,
the fixed point moves right by $\left(2*0.001\right)^{1/3}=0.12549$,
and the slope at the fixed point gives a relaxation time of 42 iterations.
Right, a much larger input of 0.11 gives rise to a displacement of
0.55, with a relaxation time of 1.8 iterations. \label{fig:fixpointdiagonal}}

\end{figure}
A compressive power law amplifies small values and attenuates large
values, because the derivative of $x^{\alpha}$ for $\alpha<1$ diverges
as $x\to0$ and conversely approaches 0 as $x\to\infty$ , hence the
name \emph{compression}. One classic example is Galileo's observation
of relief in the Moon, as the glancing incidence of light at the terminator
line amplifies relief; if the relief is about $h$ in height and the
terminator line wanders $W$ from the median, then $W=\sqrt{2hR}$,
so if the relief is $5km$ in height, then $W=\sqrt{2\times5km\times1740km}\approx132km$.

\begin{figure}[tbh]
\includegraphics[width=9cm]{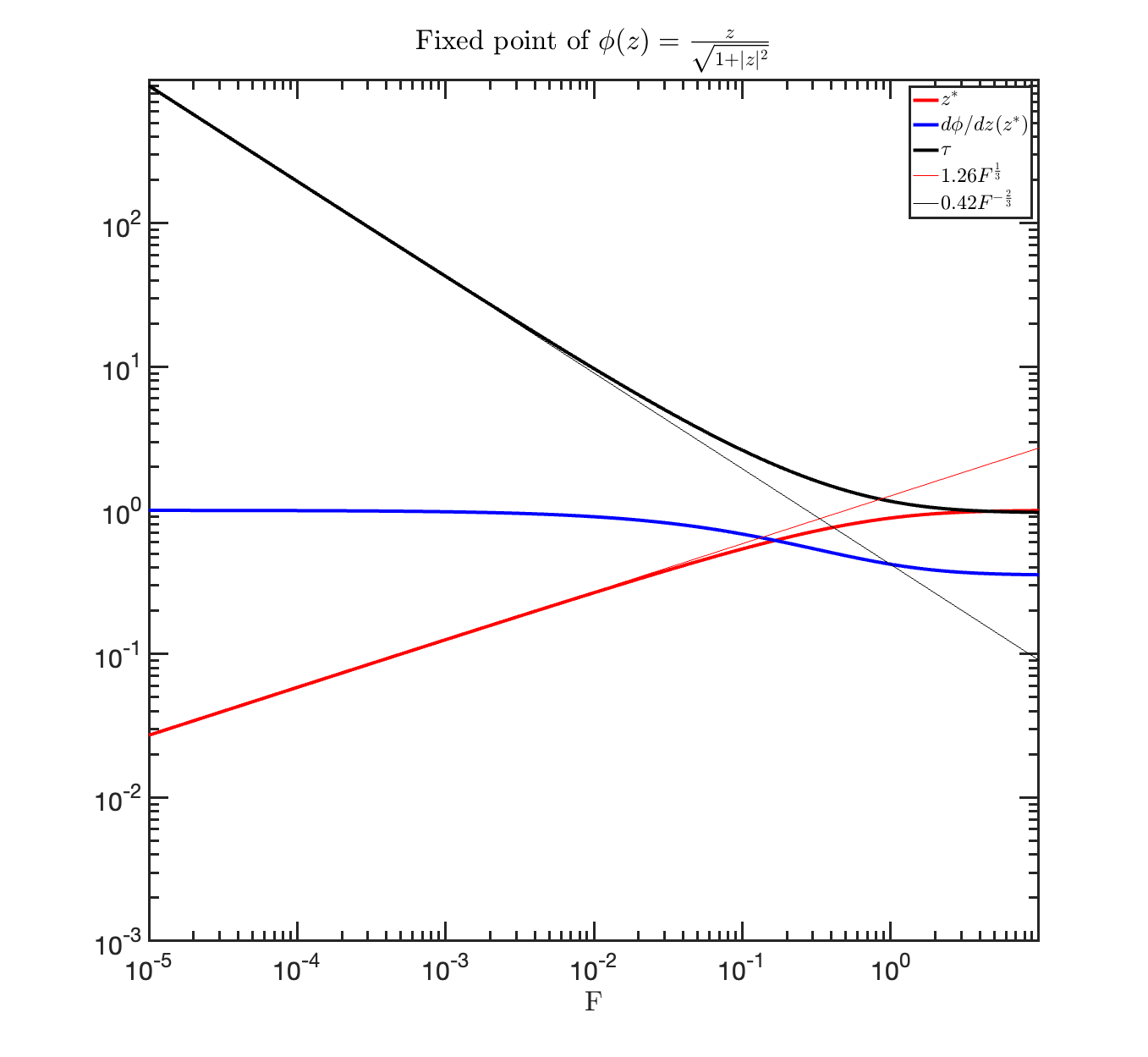}\caption{Fixed point and slope at the fixed point, as a function of input.
\label{fig:Fixpoint}}
\end{figure}

Critical systems display powerful compressive nonlinearities; for
example, a system poised exactly at a Hopf bifurcation, while forced
exactly at the resonant frequency, displays a cubic root response
\citep{essential}. Naturally this behavior extends to our system.
Consider the 1D recurrence
\begin{equation}
z_{n+1}=\phi(z_{n}+I)=\frac{z_{n}+I}{\sqrt{1+(z_{n}+I)^{2}}}\label{eq:onedrecurrence}
\end{equation}
Because all slopes are $\le1$ this recurrence is a contraction, and
has a single fixed point $z^{*}=\phi(z^{*}+1)$, which leads to 
\[
I=\phi^{-1}(z^{*})-z^{*}
\]
where $\phi^{-1}(z)=z/\sqrt{1-z^{2}}$. 

Because the slope of $\phi$ at the origin is 1, and therefore is
tangent to the diagonal line, adding a tiny $I$ will displace the
fixed point by a considerable amount, as illustrated in Figure \ref{fig:fixpointdiagonal};
in fact proportional to $(2I)^{\frac{1}{3}}$ for small enough $I$,
as can be seen by expanding 
\[
I=\frac{z^{*}}{\sqrt{1-z^{*2}}}-z^{*}\approx\frac{z^{*3}}{2}+\frac{3z^{*5}}{8}+\frac{5z^{*7}}{16}+O(z^{*9})
\]
where for sufficiently small $I$ then 
\begin{equation}
z^{*}\approx\left(2I\right)^{\frac{1}{3}}\label{eq:cubicroot}
\end{equation}
The slope at the fixed point then is $\phi'(z^{*}+I)=\left(1+(z^{*}+I)^{2}\right)^{-3/2}$;
the decay time $\tau$ is defined by $\phi'^{\tau}=e^{-1}$ or $\tau=-1/\log(\phi')$,
and with $\log\left(\left(1+(z^{*}+I)^{2}\right)^{-3/2}\right)\approx-\frac{3}{2}\left(z^{*}+I\right)^{2}\approx-\frac{3}{2}\left(2I\right)^{\frac{2}{3}}$
then we get 
\begin{equation}
\tau\approx\frac{2}{3}(2I)^{-\frac{2}{3}}\label{eq:decaytime}
\end{equation}
This equation then defines that \emph{the timescale of relaxation
in this system is a function of the input. }See Figure \ref{fig:Fixpoint}. 

\subsection{The eigenvalues of $U\otimes$}

As described above in Notation, the representation of $\left[K\otimes\right]$
is a matrix and thus it has eigenvalues and eigenvectors. We now will
show exactly what these are. 

The Fourier Convolution theorem states that the Fourier transform
of a convolution is the point-wise product of the Fourier transforms.
In our specific case, it means that we can FFT the layer $Z$ and
the kernel $K$ and element-wise multiply them. Abusing our notation
yet again, 
\[
\left[K\otimes Z\right]_{\mathcal{\mathscr{F}}}=\left[K\right]_{\mathscr{F}}.*\left[Z\right]_{\mathscr{F}}
\]
where the $.*$ operator is point-wise multiplication of two vectors
and the subscript means ``in the Fourier basis''. How can we reconcile
this with 
\[
\left[K\otimes Z\right]_{\mathscr{F}}=\left[K\otimes\right]_{\mathscr{F}}\times\left[Z\right]_{\mathscr{F}}
\]
where $\left[K\otimes\right]_{\mathscr{F}}$ is a matrix and $\times$
is matrix multiplication? The way to promote point-wise multiplication
by a vector to matrix multiplication is if the matrix is diagonal: 

\[
\left[K\otimes\right]_{\mathscr{F}}={\rm diag}\left(\left[K\right]_{\mathscr{F}}\right)
\]
where the diag operation is to embed a vector as the main diagonal
of a matrix leaving all other elements zero. 

Therefore \emph{the elements of the Fourier transform of the kernel
$K$ are the eigenvalues of the matrix $\left[K\otimes\right]$ and
thus of the kernel $K$ itself. }And, since the representation $\left[K\otimes\right]_{\mathscr{F}}$
is diagonal in the Fourier basis, the eigenvectors of $K$ are \emph{the
Fourier basis functions themselves, }namely $e^{2\pi i\frac{kj}{n}}$
etc. This is a direct result of the translational invariance of the
convolution operation. 

All of the computational advantages of the present model are given
by the ability to diagonalize (an $N^{3}$ operation in general) by
Fast Fourier transformation (an $N\ln N$ operation). Evidently, since
any kernel $K$ is diagonal on the same Fourier basis, all convolutions
commute as they share eigensystems. 

\subsection{Perturbation analysis using a unitary $U$}

A small perturbation to the initial condition $Z_{0}$ propagates
forward using the chain rule\citep{strogatz}
\[
\frac{\partial Z_{n}}{\partial Z_{0}}=\prod_{i=0}^{n-1}\Lambda_{i}\left[U\otimes\right]
\]
where the $\Lambda_{i}$ are diagonal matrices whose elements are
\[
\Lambda_{i}=\mathrm{diag}\left(\left.\frac{\partial\phi}{\partial x}\right|_{x_{i}}\right)=\mathrm{diag}(\phi'(U\otimes Z_{i}+I_{i}))
\]
For small values of the activities, the elements of $\Lambda$ are
close to 1, and so the evolution is determined by the eigenvalues
of $[K\otimes]$ and whether their powers diverge or go to zero is
determined exclusively by their absolute values. It is therefore of
interest to use kernels all of whose eigenvalues have unit absolute
value. Generating unitary matrices, otherwise known as the Stieffel
manifold parametrization problem, is computationally expensive in
the general case, as the simplest method is through exponentiation
of an anti-Hermitian matrix. On the other hand as shown above, unitary
kernels can be generated by using the convolutional exponential, which
is fast because kernels are diagonal in the Fourier basis and so FFT
can be used. 

\subsection{Dispersion relation}

The FFT of the kernel gives the resonance frequencies, so plotting
these as a function of $\left(k_{x},k_{y}\right)$ literally draws
out the dispersion relation. 

We will examine some number of kernels, including the standard $i\Delta$,
the diffractive coupling or free Schrödinger equation, plus various
couplings with random elements. One specific random kernel we shall
come back to using consists of random numbers within a disk of radius
$R$ and zero outside. This allows the kernel to be somewhat local,
and the frequencies to vary continuously in $\left(k_{x},k_{y}\right)$
space. 

Finally note the most important distinction is whether the antiHermitian
generator for the kernel $K$ is purely imaginary (symmetric) or purely
real (antisymmetric) or full antiHermitian (a linear combination of
both). 

\section{Results}

\subsection{Resonance and nonlinear compression}

\begin{figure}[tbh]
\includegraphics[width=8.4cm]{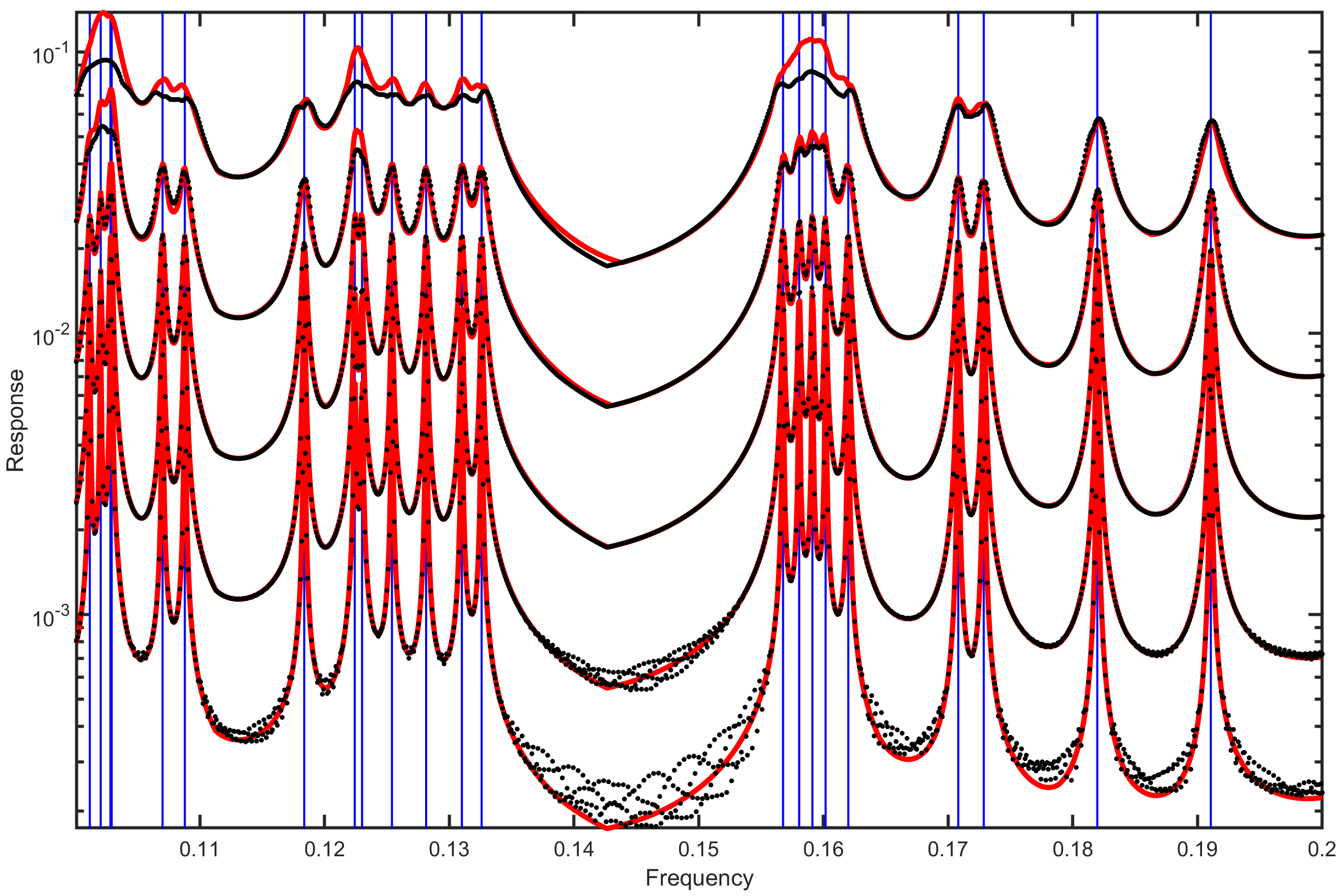}

\includegraphics[width=8.4cm]{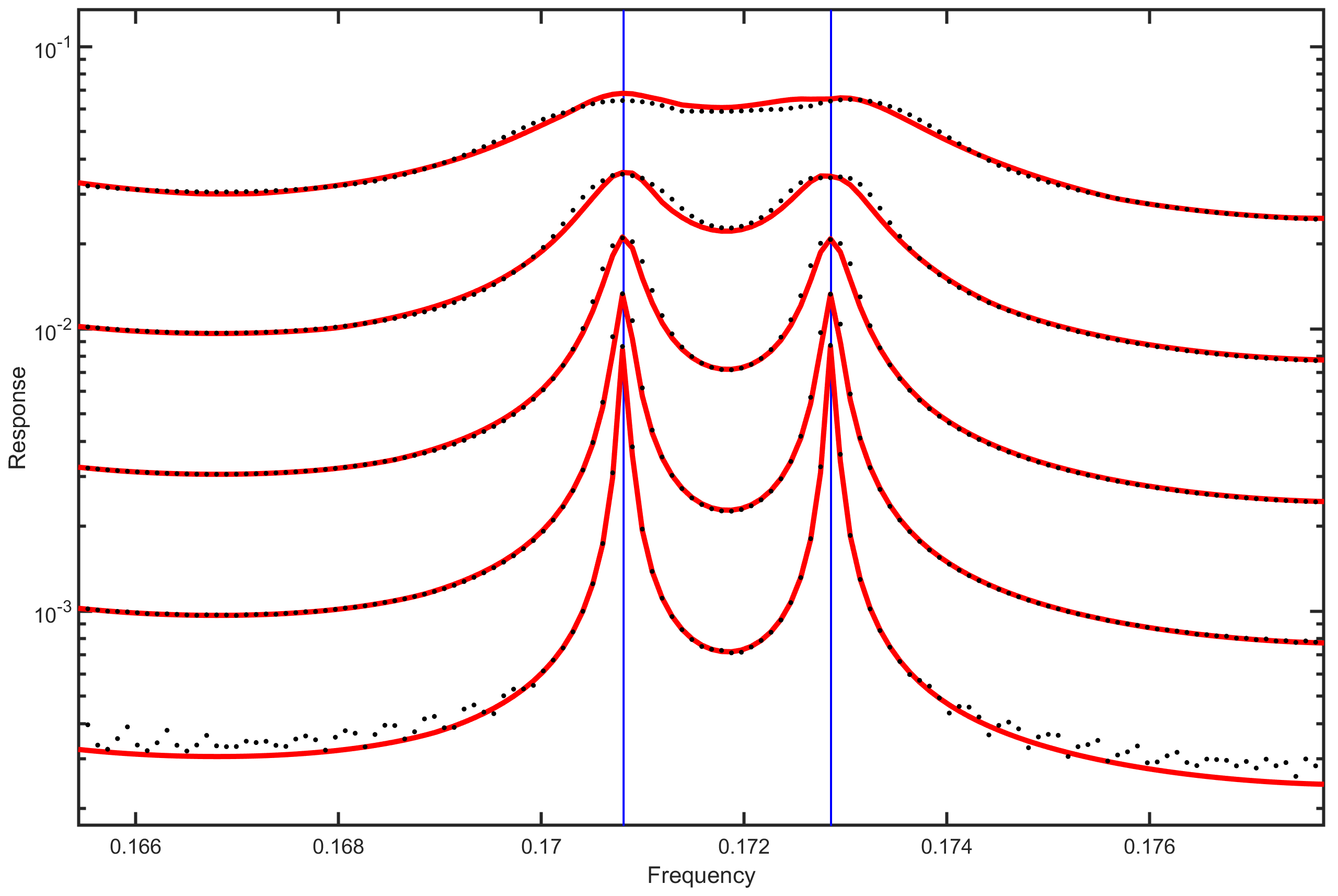}

\caption{The eigenvalues of $K$ are resonance frequencies of the system, in
that forcing at exactly that frequency causes the system to give a
disproportionately large response. Resonances of the system show compressive
nonlinearities at the resonances, linear behavior away from the resonances,
and broadening of the responses as the forcing amplitude increases
to match both regimes to each other. Shown are the amplitude of the
responses to 5 different intensities separated by $\sqrt{10}$ from
each other. Black dots are the result of numerical simulation (N=1024,
compleetly random $K$, $2^{24}$ iterations, 1024 frequencies, 5
intensities $10^{-(1:5)/2}$). Red lines are the analytical calculation
of Eq. \ref{eq:nonperturb}. The blue lines are the eigenvalues of
the kernel. Bottom panel: detail showing the widening as the amplitude
of the input increases. \label{fig:resonances}}

\end{figure}

Consider forcing the system \ref{eq:model} with a periodic force
of amplitude $\alpha$ and frequency $e^{i\theta}$ at one single
site, e.g. the origin: 
\begin{equation}
I_{n}=\alpha e^{in\theta}\delta_{0}\label{eq:singlesiteperiodic}
\end{equation}
where $\delta_{0}$ is an array the same size as the layer with a
single 1 at site 0 and zero otherwise. When $\alpha\ll1$ and the
frequency $e^{i\theta}$ coincides exactly with an eigenvalue of $K$,
then the system resonates and the growth is only limited by reaching
the nonlinear regime of $\phi.$ These resonances show the characteristic
$1/3$-power law scaling and width broadening described in detail
in \citep{essential}. See Figure \ref{fig:resonances} 

We will revisit this subject later in order to derive an analytical
solution to this diagram. 

\subsection{Background-dependent wave attenuation}

We now consider an input consisting of a fixed constant background
$I^{*}$ plus a small input $I_{n}=I^{*}+S_{n}$ where $S_{n}\ll1$.
We follow standard dynamical systems perturbation theory

\[
Z_{n+1}=\phi(U\otimes Z_{n}+I_{n})
\]
we compute the fixed point for the input $I^{*}$ alone
\[
Z^{*}=\phi(U\otimes Z^{*}+I^{*})
\]
and since a constant input is always an eigenvector of $[U\otimes]$,
calling its eigenvalue $\lambda_{0}$, we get $Z^{*}=\phi(\lambda_{0}Z^{*}+I^{*})$
which is now a scalar equation. We will then denote the response to
$I_{n}=I^{*}+S_{n}$ as $Z_{n}=Z^{*}+R_{n}$, and substituting
\[
Z^{*}+R_{n+1}=\phi(U\otimes Z^{*}+U\otimes R_{n}+I^{*}+S_{n})
\]
and assuming that $R_{n}$ is also small (not true at resonance) we
can expand to 1st order to get
\[
Z^{*}+R_{n+1}=\phi(\lambda_{0}Z^{*}+I^{*})+\phi'(\lambda_{0}Z^{*}+I^{*})(U\otimes R_{n}+S_{n})
\]
where the constant slope of the activation function at the fixed point
\begin{equation}
\gamma\doteq\phi'(\lambda_{0}Z^{*}+I^{*})\label{eq:gamma}
\end{equation}
will be called $\gamma$, from where we get the linear equation
\[
R_{n+1}=\gamma(U\otimes R_{n}+S_{n})
\]
Since this equation is diagonal in Fourier space, using tildes to
denote the (spatial) Fourier transforms (e.g. $\tilde{R}=\mathscr{F}[R]$)
we get 
\begin{equation}
\tilde{R}_{n+1}=\gamma\tilde{U}\tilde{R}_{n}+\gamma\tilde{S_{n}}\label{eq:perturb}
\end{equation}
where the product between the Fourier transforms of the kernel and
the perturbation is again \emph{element-wise}. Remembering that the
Fourier transform of the kernel consists of its eigenvalues as a linear
operator and that, in the case we have hitherto considered in which
$U=e_{\otimes}^{A}$ is the convolutional exponential of an anti-Hermitian
kernel. Evidently $|\gamma|<1$ is needed for a stable solution to
this perturbative equation. 

There is a slight issue though: $\phi$ is not an analytic function,
and thus its derivative is not the same in every direction. The derivative
of $\phi$ is $\gamma$ along the radial direction, but along the
angular direction Eq. \ref{eq:rotinv} prescribes the derivative to
be $i/\sqrt{(1+x^{2}}$. As the argument continuously rotates, an
effective $\gamma$ ensues, which is $\gamma_{{\rm eff}}=\left(1+2\gamma\right)/3$. 

\subsection{Attenuation of periodic single-site forcing}

\begin{figure}
\includegraphics[width=8.4cm]{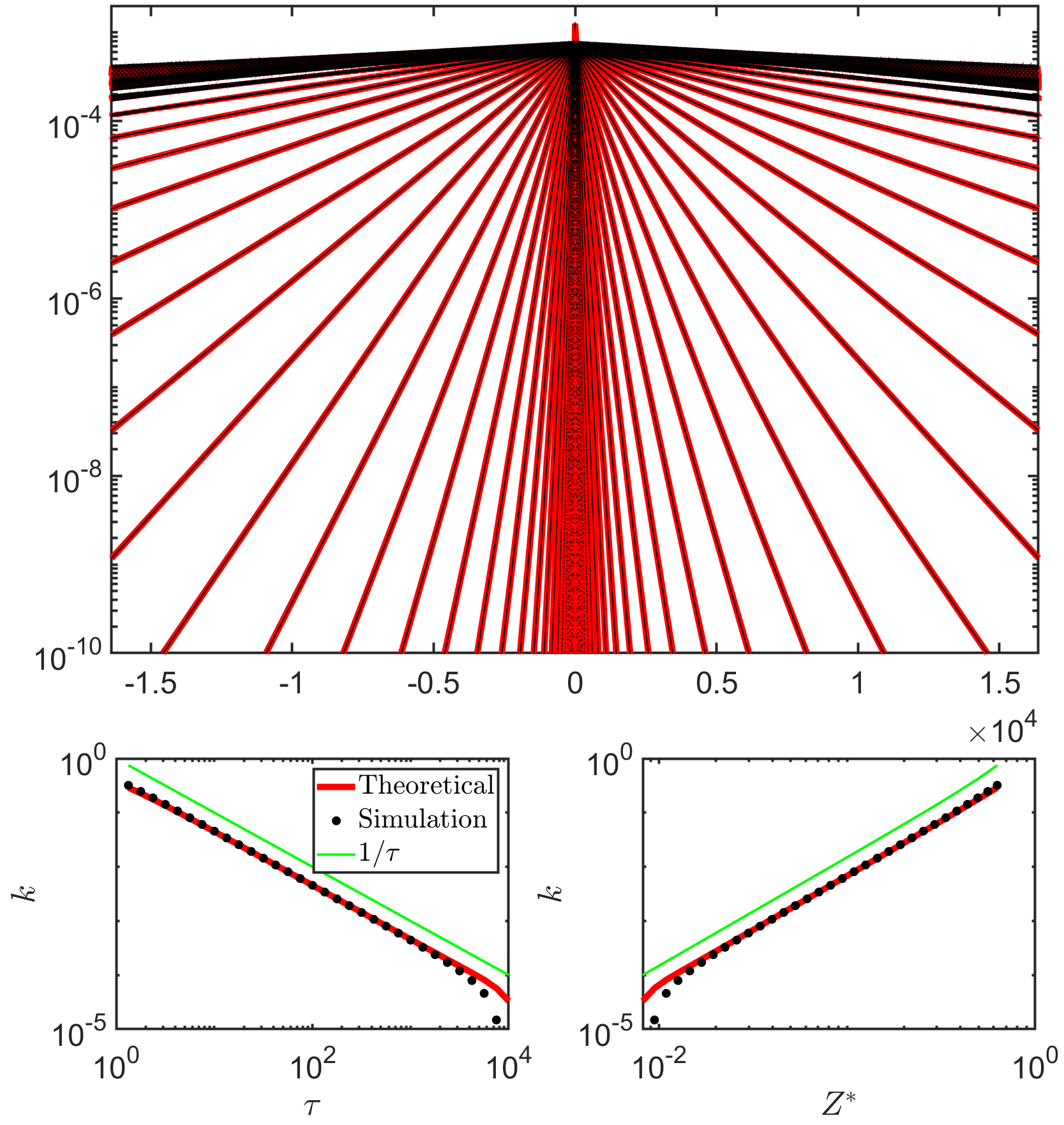}\caption{Attenuation of a propagating wave. Top panel, an oscillatory forcing
is injected at the center, and propagates outwards, attenuating as
it does. In the asymptotic state, away from the injection site $R^{*}\approx e^{-\lambda|x|}$.
The different lines were generated by choosing $\tau=10^{(1:32)/8}$
and generating the corresponding $I^{*}$. $N=2^{16}$, $\alpha=10^{-3}$,
$U=e_{\otimes}^{i\Delta}$, $\lambda=e^{-i}$. Black dots, full numerical
simulation, $2^{17}$ timesteps. Red solid curves, from Eq. \ref{eq:analyticperturbation}.
Bottom left, $\lambda$ vs. $\tau$ show an inverse relationship.
Green line is $1/\tau$. Bottom right, direct power law between $\lambda$
and $Z^{*}.$\label{fig:attenuation}}

\end{figure}

\begin{figure}

\includegraphics[width=8.4cm]{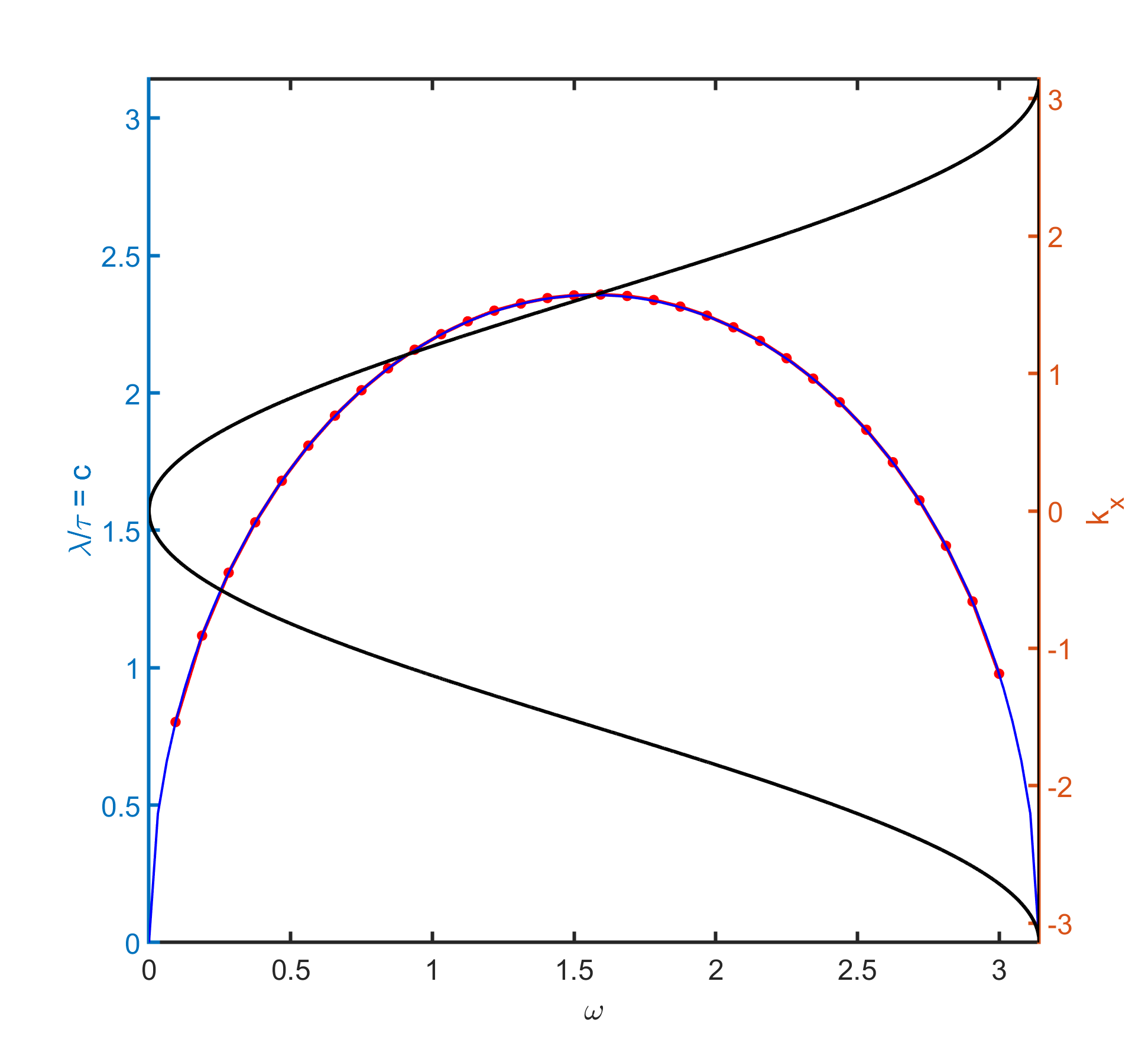}\caption{As shown in the prior figure, the decay lengthscale $\lambda$ is
proportional to $1/\tau$; the value of $1/(\lambda\tau)$ has units
of space/time and is a speed $c_{eff}$. For multiple values of $\omega$
and $\tau$ the propagation solutions were computed, and linear fits
to the log of $|R^{*}|$ were used to evaluate $\lambda$. At any
given frequency, $c_{eff}$ was estimated as the median of $1/(\lambda\tau)$
for that frequency and all $\tau.$ Left axis, $c_{eff}$ vs. the
forcing frequency (red), and $c_{g}\protect\doteq\partial\omega/\partial k=\frac{3}{2}\sqrt{\omega(\pi-\omega)}$
(blue). Right axis, rotated dispersion relation for the $U=e_{\otimes}^{i\Delta}$,
where it can be seen that $\omega=\frac{\pi}{2}(\cos(k)-1)$. \label{fig:groupvelocity}}
\end{figure}

Consider the sub-case in which $S_{n}$ is applied at a single site
(wlog the origin) and is a periodic function of time; $S_{n}=\alpha e^{in\theta}\delta_{j0}$
where $j$ is an index into the layer's individual elements, and we
expect that asymptotically the response will also oscillate like $R_{n}\approx R^{*}e^{in\theta}$
. The spatial Fourier transform of $\delta_{i0}$ is a constant vector.
Substituting 
\[
\tilde{R^{*}}e^{in\theta}e^{i\theta}=\gamma_{{\rm eff}}\tilde{U}\tilde{R^{*}}e^{in\theta}+\gamma_{{\rm eff}}\alpha e^{in\theta}\tilde{\delta}
\]
where again multiplication is element-wise element-wise leads to the
solution 

\[
\tilde{R^{*}}=\cdot\frac{\gamma\alpha\mathbf{1}}{\left(e^{i\theta}-\gamma_{{\rm eff}}\tilde{U}\right)}
\]
where the period in front of the division sign reminds us again that
this division is \emph{element-wise}. Inverting the FFT 
\begin{equation}
R^{*}=\mathscr{F}^{-1}\left[\cdot\frac{\alpha\mathbf{1}}{e^{i\theta}/\gamma_{{\rm eff}}-\tilde{U}}\right]\label{eq:analyticperturbation}
\end{equation}
The closer $e^{i\theta}$ is to an eigenvalue of $K$ and the closer
$\gamma$ is to 1 we approach a resonant situation. For kernels with
a smooth dispersion relation, like a Laplacian kernel, $R^{*}$ decays
exponentially away from the site of injection like $e^{-\lambda|x|}$;
the spatial decay constant $\lambda$ is inversely proportional to
the temporal decay constant $\tau\equiv-1/\log\gamma$ through the
group velocity $\partial\omega/\partial k$, or otherwise the density
of eigenvalues around the forcing frequency. Please see Figure \ref{fig:attenuation},
where both full numerical simulations as well as this analytical solution
for various intensities of $I^{*}$ and thus of $\gamma$. Fitting
straight lines to the flanks of $R^{*}$ to obtain $\lambda$ we observe
that indeed $\lambda\approx1/\tau$. In Figure \ref{fig:groupvelocity},
the value of $c_{eff}\doteq1/(\lambda\tau)$ is computed for various
different forcing frequencies, and shown to agree exactly with the
value of the group velocity $c_{g}\doteq\partial\omega/\partial k$,
which for this particular kernel equals $\frac{3}{2}\sqrt{\omega(\pi-\omega)}$. 

\subsection{Periodic forcing at one single element, analytic solution}

Equipped with Equation \ref{eq:analyticperturbation}, we can now
attack again the problem of the responses to a single frequency at
a single point, but with no background activity. As described above,
the response will grow in time until it saturates, and it will saturate
at a level that is nonlinearly dependent on the amplitude of the incoming
oscillation through a compressive nonlinearity of order $1/3$. In
steady state, 
\[
R^{*}e^{in\theta}e^{i\theta}=\phi\left(U\otimes R^{*}e^{in\theta}+\alpha e^{in\theta}\delta_{i0}\right)
\]

We will now investigate whether these steady state solutions conform
to \ref{eq:analyticperturbation}; by inverting Equation \ref{eq:analyticperturbation},
we get a simple diagnostic for whether this solution obeys this expression: 

\[
\left(\cdot\frac{\alpha}{\mathscr{F}\left[R^{*}\right]}+\tilde{U}\right)e^{-i\theta}=\gamma^{-1}
\]
or, namely, every single element of the expression on the left should
be equal to $1+\epsilon$ with $\epsilon\ll1$. Numerical simulation
shows this is indeed the case, although it takes an extremely long
time to converge for most elements when the $\alpha$ is very small.
However, the element closest to the driving frequency (the pole) converges
rather rapidly because it is largest in amplitude; furthermore it
shows that $\epsilon,$ which measures the distance to the unit circle,
is a power law: $\epsilon=PN^{-2/3}\alpha^{2/3}$, with the proportionality
constant $P$ being of order $1$ and dependent only on the degeneracy
of the eigenvalue (i.e. in 1D $P=(1/2)^{1/3}$ for non-degenerate
eigenvalues, or $P=(3/2)^{1/3}$ for a purely imaginary symmetric
$aH$, where the eigenvalues are two-fold degenerate)

Therefore we can reconstruct the solution using Equation \ref{eq:analyticperturbation}
and using
\[
\gamma^{-1}=1+P\left(\frac{\alpha}{N}\right)^{\frac{2}{3}}
\]
namely 

\begin{equation}
R^{*}=\mathscr{F}^{-1}\left[\cdot\frac{\alpha\mathbf{1}}{\left(1+P\left(\frac{\alpha}{N}\right)^{\frac{2}{3}}\right)e^{i\theta}-\tilde{K}}\right]\label{eq:nonperturb}
\end{equation}
Figure \ref{fig:resonances} shows the close agreement between this
solution and the numerical simulations. The analytic solution deviates
slightly from the numerical calculations at the largest values of
$\alpha$ and exactly on the resonances, where the amplitude of the
solution no longer follows the power law. Elsewhere the agreement
is full. 

\section{Conclusions}

If you raise the cover of a piano, depress the right pedal to lift
the felt dampers from the strings, and then loudly yell into the back,
you'll hear the strings resonate sympathetically with the components
of your voice; for several seconds after your vocalization you'll
hear a memory of the timbral characteristics of it. If you yell at
a different pitch it will keep a different memory. This classical
demonstration of resonance and spectral decomposition is, in its own
way, a demonstration of memory. As a dynamical mechanism it is diametrically
opposite to the Hopfield network; reverberation stores information
in center manifolds as opposed to fixed points. 

However the purely resonant memory does not explicitly retain the
passage of time; for that we need \emph{propagation }in addition to
resonance, so that a position in space encodes time elapsed. The cuRNNs
discussed in this Paper provide such propagation; this is more explicit
when the kernel is spatially compact, but for kernels with bigger
spatial support, information is still \emph{reversibly} scrambled
as time elapses, by scrambling phases of oscillations. Recent studies
have found a central role to wave propagation in both instantiating
a ``working memory'' of the RNN \citep{waves1} as well as transferring
information between working variables as well as storing it \citep{waves2}.
These traveling waves are naturally associated to eigenvalues of the
synaptic matrix of unit absolute value. It is considerably easier
to study such traveling waves in a convolutional system, as the fast
convolutional exponential allows us both to generate the unit eigenvalues
that support such waves, as well as relating those eigenvalues directly
to the eigenvector that represents its support in the units. 

The center concern of this Paper has been the modulation of spatial
and temporal scales by the action of the input. In a critical system,
there are no natural timescales or lengthscales, both being formally
infinite; external inputs then lower these to finite values through
interactions between the input and the nonlinear terms. Input-dependent
changes in spatial propagation and relaxation timescales have been
experimentally observed e.g. in primary visual cortex \citep{scale1,scale2,scale3,scale4},
and some features thereof have been found to agree with critical models
\citep{waveatten,waveatt2}, yet comprehensive fundamental theories
are still at large. We have given here a pretty complete analytic
treatement of how timescales and lengthscales of traveling waves are
dynamically modulated by ongoing activity, as a direct consequence
of the critical nature of our system. In oncoming work we shall show
how such ongoing activity can be spatially patterned to dynamically
create channels for these waves.

\end{document}